\documentclass[traditabstract]{aa} 
\usepackage{txfonts}
\usepackage{graphicx}
\usepackage{natbib}

\newcommand{\dpleo}{\mbox{\object{DP\,Leo}}}
\newcommand{\ten}[2]{#1\times 10^{#2}}
\begin{document}
\title{Evidence for an oscillation of the magnetic axis of the white dwarf in the polar DP Leonis}

\author{
Beuermann, K. \inst{1} 
\and 
Dreizler, S. \inst{1} \and 
Hessman, F.~V. \inst{1} \and 
Schwope, A.~D.\inst{2}  } 

\institute{
Institut f\"ur Astrophysik, Georg-August-Universit\"at, Friedrich-Hund-Platz 1, D-37077 G\"ottingen, Germany 
\and Leibniz-Institut f\"ur Astrophysik Potsdam (AIP), An der Sternwarte 16, 14482
Potsdam, Germany
}
\date{Received 5 December 2013; accepted 31 Dec 2013}

\authorrunning{K. Beuermann et al.} 
\titlerunning{Oscillation of the magnetic axis of the white dwarf in DP Leo}

\abstract{From 1979 to 2001, the magnetic axis of the white dwarf in
  the polar DP Leo slowly rotated by $50^\circ$ in azimuth, possibly
  indicating a small asynchronism between the rotational and orbital
  periods of the magnetic white dwarf. Using the MONET/North
  telescope, we have obtained phase-resolved orbital light curves
  between 2009 and 2013, which show that this trend has not continued
  in recent years. Our data are consistent with the theoretically
  predicted oscillation of the magnetic axis of the white dwarf about
  an equilibrium orientation, which is defined by the competition
  between the accretion torque and the magnetostatic interaction of
  the primary and secondary star. Our data indicate an oscillation
  period of $\sim\!60$\,yr, an amplitude of about 25\degr, and an
  equilibrium orientation leading the connecting line of the two stars
  by about 7\degr.}

\keywords{ Stars: binaries: close -- Stars: binaries: eclipsing --
  Stars: white dwarfs -- Stars: Cataclysmic Variables -- Stars:
  individual: \dpleo }

\maketitle


\section{Introduction}

Since its discovery by \citet{biermannetal85}, the 18 mag short-period
magnetic cataclysmic variable or polar \dpleo\
($P_\mathrm{orb}\!=\!89.8$\,min) has shown an accretion geometry,
characterized by a prime accreting pole that points approximately
toward the secondary and a weakly accreting pole in the opposite
hemisphere. The former is responsible for the optical cyclotron and
X-ray emission in the `bright phase', which lasts slightly longer than
half the orbital period
\citep{schaafetal87,baileyetal93,robinsoncordova94,pandeletal02,schwopeetal02}.
The latter usually emits only faint X-ray and cyclotron
radiation with prominent cyclotron emission lines. The magnetic field
strengths are 30.5\,MG and 59\,MG in the accretion regions near the prime
and second pole, respectively \citep{cropperwickramasinghe94}.

The mechanism responsible for the attainment of synchronism in polars,
that is the equality of rotational and orbital periods of the white
dwarf, is thought to be the interaction between the magnetic moments
of primary and secondary. Synchronism is attained if the magnetic
torque dominates the accretion torque. A variable accretion torque or
some other perturbation may lead to an oscillation (or libration) of
the magnetic axis about its equilibrium position, measured as a
secular variation of the longitude of the accretion spot. The
predicted period of such an oscillation is of the order of 50\,yr
\citep{kingetal90,kingwhitehurst91,wickramasinghewu91,wuwickramasinghe93,campbellschwope99}. If
equilibrium is not attained, the white dwarf may rotate slowly under
the action of the accretion torque. Four polars lack strict
synchronization, displaying a difference between rotational and
orbital periods on the order of 1\% \citep{campbellschwope99}.

The timing observations of \dpleo\ available until 2001 were
summarized by \citet{pandeletal02} and \citet{schwopeetal02}. They
noted that between 1979 and 2001, the longitude of the accretion spot
increased linearly by about 2.3\degr\ per year, implying a rotation
period shorter than the orbital period by 5.9\,ms. At the same time,
the $O-C$ variations of the binary were interpreted as a monotonic
decrease of the orbital period by $\ten{5.3}{-12}$\,s\,s$^{-1}$,
suggesting that the binary was heading toward synchronism on a time
scale of $\sim\!30$\,yr. This interpretation appeared plausible since
the polar V1500\,Cyg displayed a similar behavior after having
suffered an outburst as Nova Cyg 1975, in which it lost synchronism
\citep{schmidtstockman91}. A decade later, however, \citet{qianetal10}
and \citet{beuermannetal11} demonstrated that a monotonous period
decrease in \dpleo\ does not exist and re-interpreted the $O-C$
variations as a light travel time effect caused by a giant planet
orbiting the binary, while the binary period remained constant. This
again raises the question of the origin of the secular variation of
the spot longitude.

In this paper, we report new measurements which show that the progression
of the spot longitude has halted since 2009. Our data are
consistent with an oscillation of the magnetic axis of the white dwarf
around its equilibrium position, placing the interpretation of the \dpleo\ data
on a new basis.

\section{Observations}
\label{sec:obs}

We observed the cataclysmic variable \dpleo\ in 45 nights between
March 2009 and March 2013, using the 1.2-m MONET/North telescope at
the University of Texas McDonald Observatory via the MONET
browser-based remote-observing interface. Photometric light curves in
white light were obtained, using an Apogee ALTA E47+ 1k$\times$1k CCD
camera. Exposure times of 30\,s or 60\,s were use for the orbital
light curves, the narrow eclipses were covered with 10\,s exposures,
all separated by 3\,s readout. Photometry was performed relative to
the r=16.86 comparison star SDSS\,J111725.54+175614.9, which is
located 2\,\farcm29 E and 1\,\farcm45 S of the binary and has colors
that approximate those of \dpleo. The images were corrected for dark
current and flatfielded in the usual way. Fluxes were extracted, using
a radius of about one FWHM of the point-spread function.

The mid-eclipse times of the white dwarf, which define orbital phase
$\phi\!=\!0$, can be determined from the observed light curves with an
accuracy better than a few seconds \citep{beuermannetal11}. The
shutter times and the time stamps of the MONET/North telescope have an
accuracy of a small fraction of a second, more than adequate for the
present purpose. We converted the measured times from UTC to
Barycentric Dynamical Time (TDB) and corrected them for the light
travel time to the solar system barycenter, using the tool provided by
\citet{eastmanetal10}\footnote{http://astroutils.astronomy.ohio-state.edu/time/}. The
linear ephemeris of \citet[][their Eq.\,1]{beuermannetal11} was found
to be still valid within a few seconds in early 2013. We conclude that
the orbital-phase error for a given measured UTC timing is less than
0.001.  The new eclipse data will be reported elsewhere.

\begin{figure}[t]
\includegraphics[bb=273 20 488 780,height=90mm,angle=-90,clip]{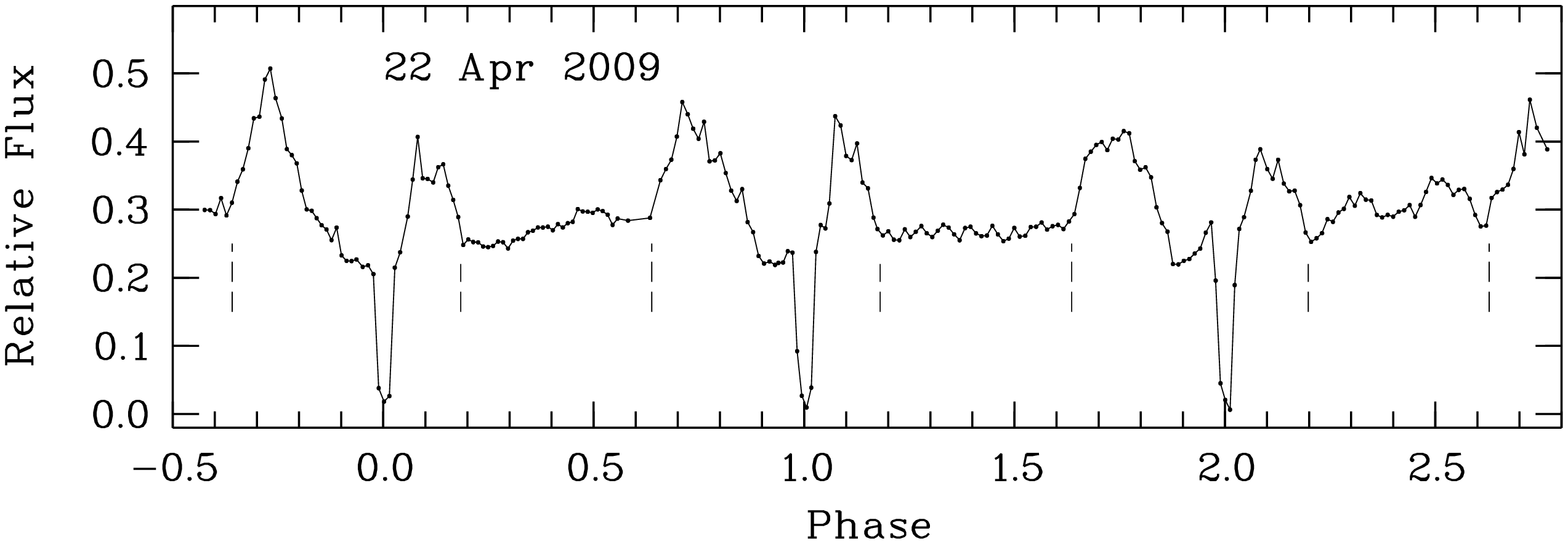}	
\includegraphics[bb=273 20 488 780,height=90mm,angle=-90,clip]{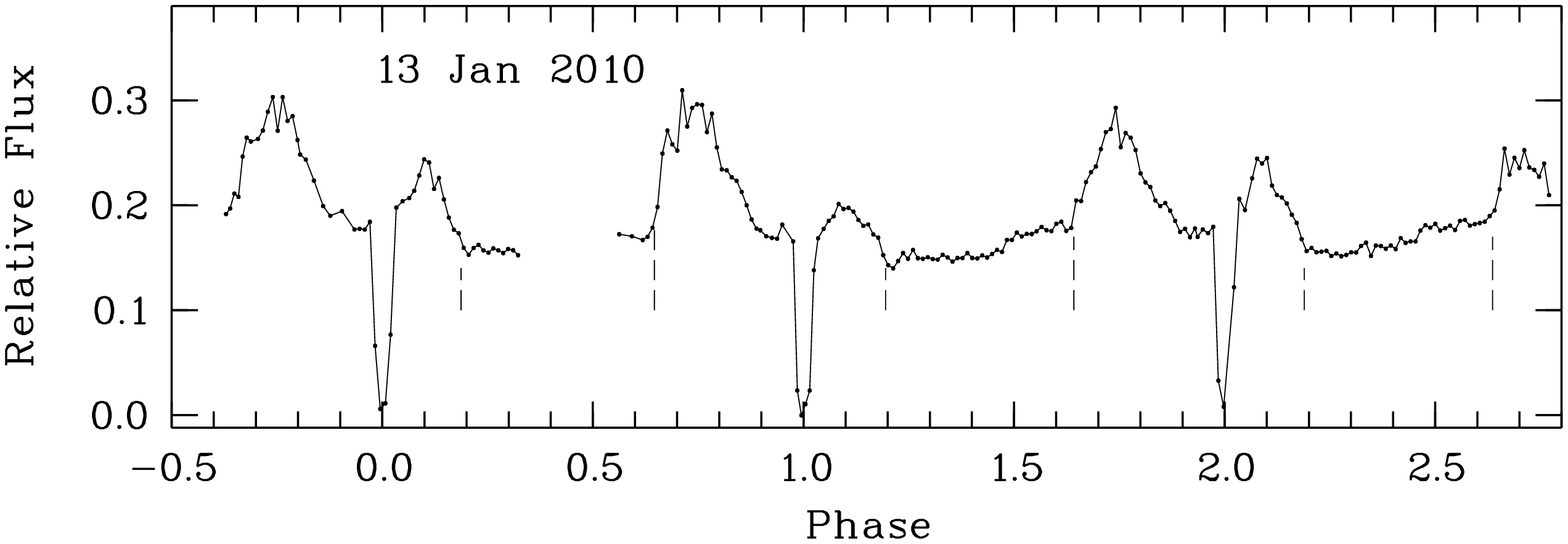}	
\includegraphics[bb=273 20 488 780,height=90mm,angle=-90,clip]{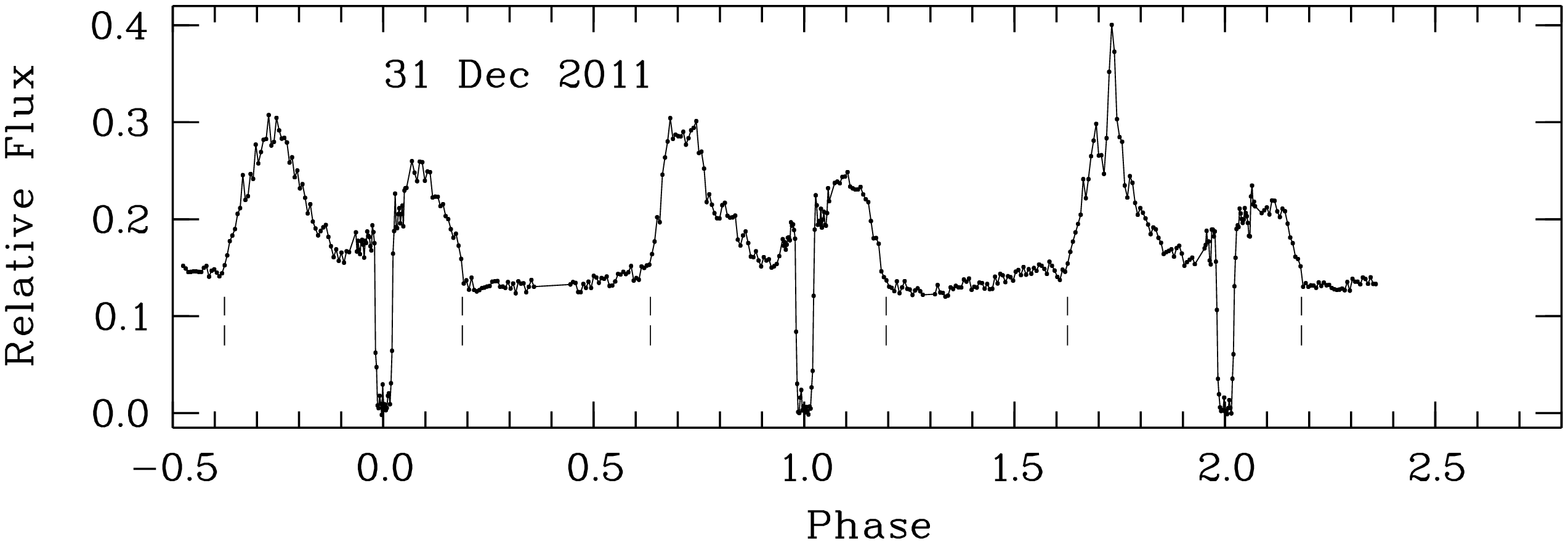}	
\includegraphics[bb=273 20 541 780,height=90mm,angle=-90,clip]{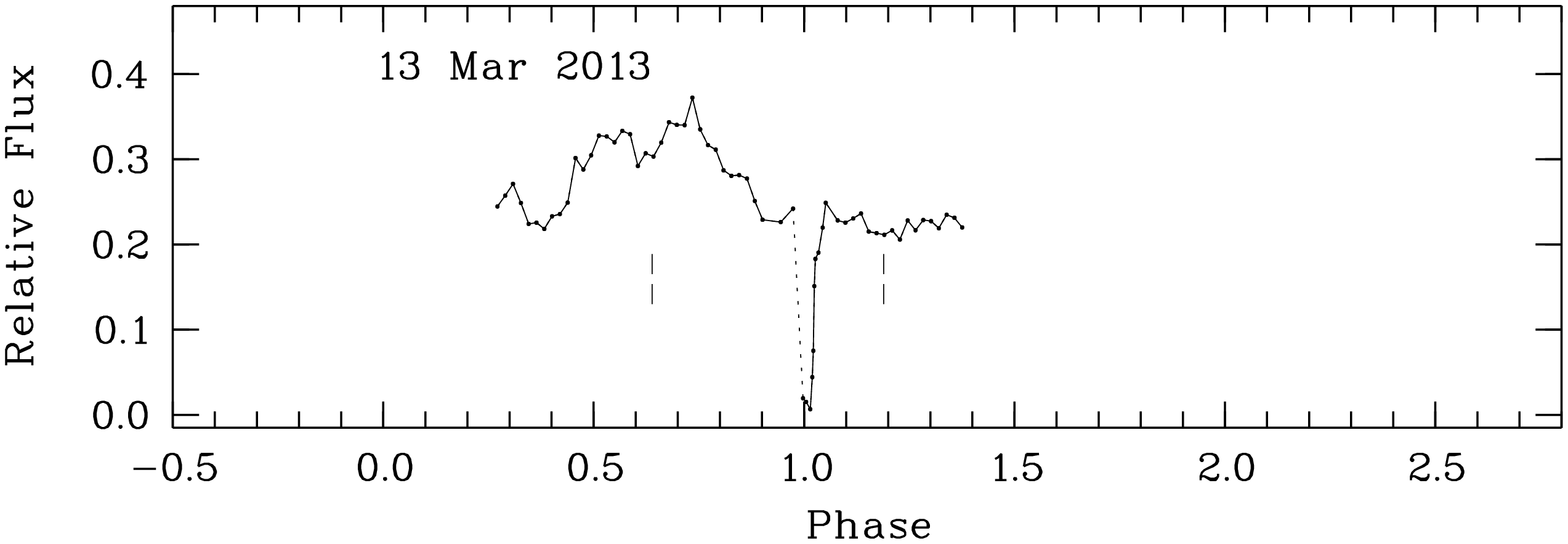}	
\caption[chart]{Examples of the light curves of \dpleo\ taken in
  white light between March 2009 and March 2013,
  phased relative to the eclipse of the white dwarf by the secondary
  star. The vertical dashed markers indicate the beginning and end of the
  visibility of the prime accretion spot (see text). }
\label{fig:lc}
\end{figure}

\section{Results}

\subsection{Orbital light curves}
\label{sec:lc}

Figure~\ref{fig:lc} shows examples of the total of 45 white-light
orbital light curves of \dpleo\ taken in five observing seasons
between March 2009 to March 2013\footnote{Spring 2009, winter/spring
  2009/2010, winter/spring 2010/2011, winter/spring 2011/2012, and
  early 2013.}. All light curves are phased relative to the
simultaneously observed eclipses of the white dwarf by the secondary
star \citep{beuermannetal11}. The general shape of the light curve is
characterized by a bright phase, which extends from about
$\phi\!\simeq\!-0.36$ to 0.19 and is caused by emission from the prime
pole visible during this interval \citep[][and references
therein]{pandeletal02,schwopeetal02}.  Cyclotron beaming is
responsible for the double-humped shape.

\begin{figure}[t]
\includegraphics[bb=60 25 562 476,width=53.5mm,angle=-90,clip]{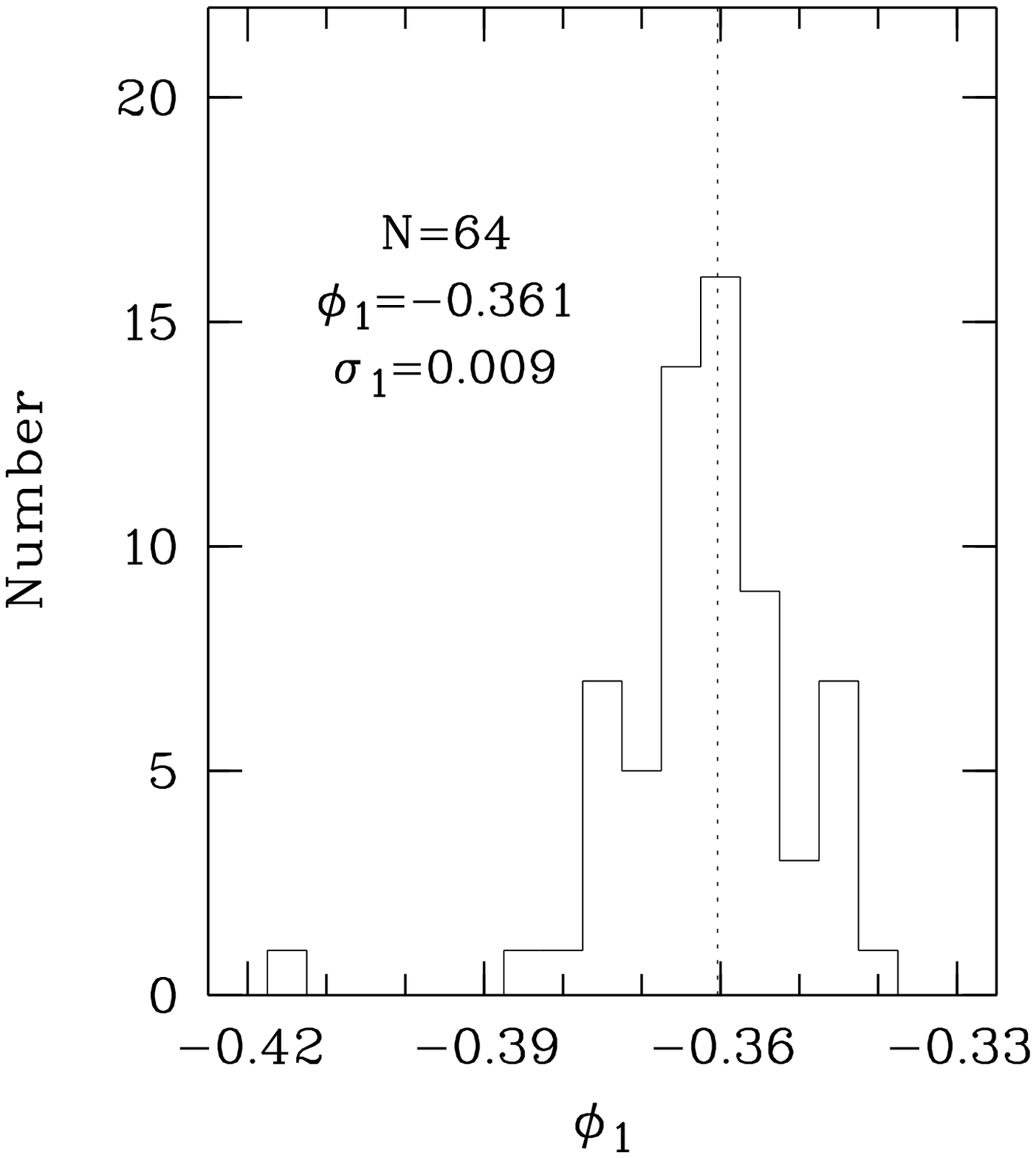}	
\includegraphics[bb=60 80 562 464,width=53.5mm,angle=-90,clip]{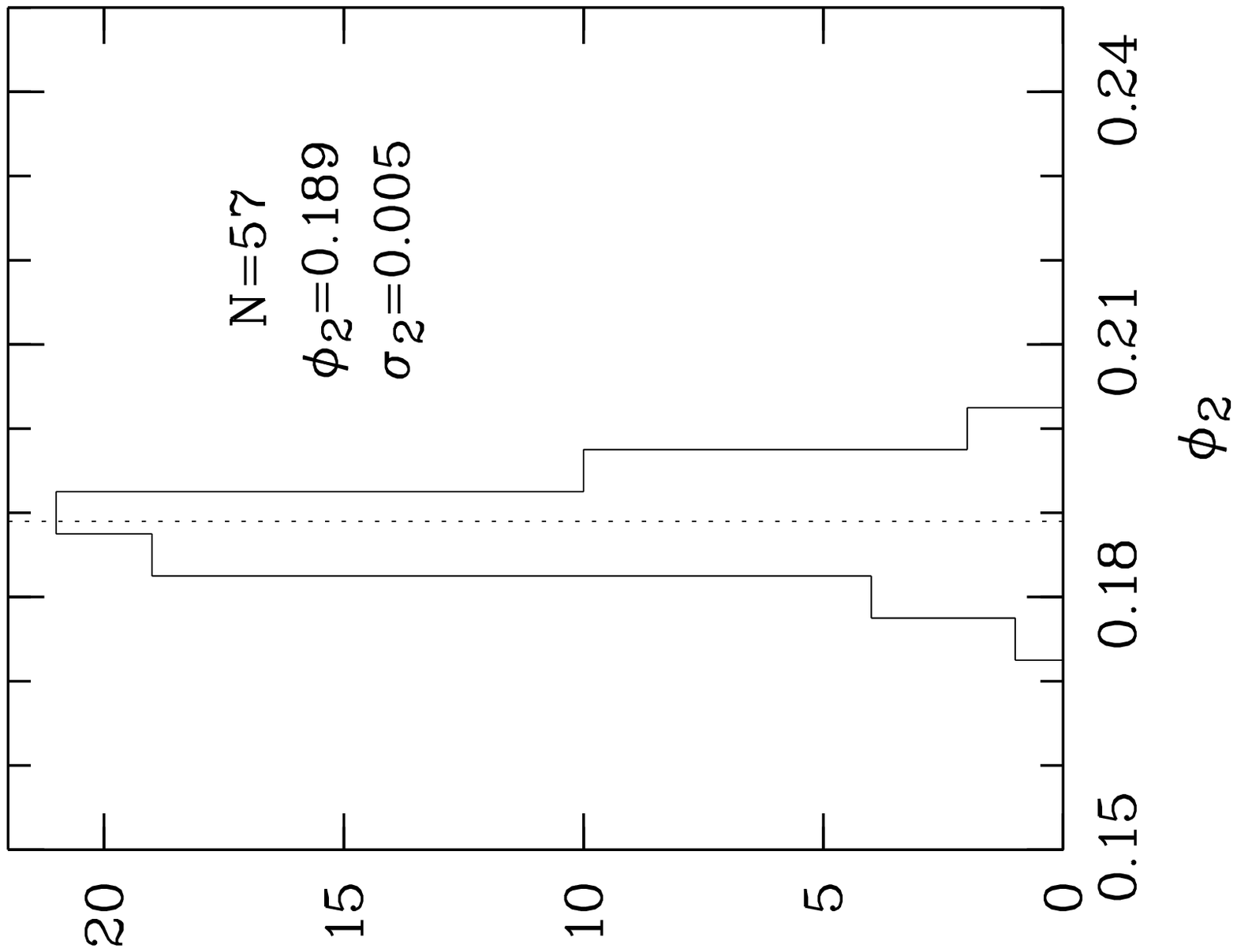}	
\caption[chart]{Distributions of the measured phases of the beginning
  (left panel) and end (right panel) of the visibility of the prime
  accretion spot. }
\label{fig:phi}

\vspace{-1mm}
\end{figure}

The flux in the phase interval, $\phi\!=\!0.19-0.64$, is cyclotron
emission from the usually fainter second pole with a small
contribution from the white dwarf photosphere (except for a low state,
when the cyclotron emission disappears). A double-humped structure is
also observed in a few instances in the faint phase. This is a
transient phenomenon, as seen in the top panel of Fig.~\ref{fig:lc},
where maxima at orbital phases of about 0.30 and 0.50 appear in the
third of three faint-phase intervals, but these maxima are absent from
the second interval. Traces appear frequently, in particular of the
$\phi\!=\!0.50$ hump. On a single occasion, on 13 March 2013, the
$\phi\!=\!0.50$ hump rivaled the prime-pole emission in brightness
(Fig.~\ref{fig:lc}, bottom panel), rendering the beginning of the
prime-pole bright phase unmeasurable.

\begin{table}[b]
\begin{flushleft}
  \caption{Longitude $\psi$ of the center of the prime-pole bright
    phase, based on $n_1$ and $n_2$ measurements of the beginning and end
    phases of the bright-pole visibility interval, $\phi_1$ and
    $\phi_2$, respectively, in five observing seasons. Phase errors
    quoted in parentheses refer to the last digits.}
\begin{tabular}{@{\hspace{0mm}}l@{\hspace{2mm}}c@{\hspace{4mm}}r@{\hspace{2mm}}c@{\hspace{2mm}}r@{\hspace{2mm}}c@{\hspace{2mm}}c}
\hline\hline \\[-1ex]
Observing & Mean JD  & $n_1$ & $\phi_1$ & $n_2$ &  $\phi_2$ & $\psi$   \\
Season    & 2450000+ &       &          &      &           & (degree) \\[0.5ex]
\hline\\[-0.5ex]
\multicolumn{7}{l}{\emph{(a) ~~Individual observing seasons}} \\[0.5ex]
2009      & 54941.7 & 13 & -0.359(8 ) & 17 & 0.187(6)  & $30.9\pm1.7$ \\
2009/10   & 55200.4 & 16 & -0.357(5)  & 14 & 0.191(3)  &  $30.0\pm1.1$   \\
2010/11   & 55624.7 & 13 & -0.356(9)  & 12 & 0.189(4) &  $29.9\pm1.7$    \\
2011/12   & 55932.6 & 14 & -0.368(8)  & 11 & 0.188(6) &  $32.3\pm1.7$ \\  
2013      & 56333.6 &  8 & \hspace{1.5mm}-0.367(12) &  3 & \hspace{1.5mm}0.189(12) &  $32.1\pm3.0$ \\ [1.3ex]  
\multicolumn{7}{l}{\emph{(b) ~~All data combined}} \\[0.5ex]
2009-2013 & 55478.2 & 64 & -0.361(9)  & 57 & 0.189(5) & $31.0\pm1.9$  \\ [1.3ex] 
\hline
\end{tabular}
\label{tab:phi}
\end{flushleft}
\end{table}

The vertical dashed markers in the top three panels of
Fig.~\ref{fig:lc} indicate the individual measurements of the
beginning and end of the visibility of the prime-pole accretion spot,
$\phi_\mathrm{1}$ and $\phi_\mathrm{2}$, respectively, in the bottom
panel they refer to the expected phases. All light curves combined
yield a total of 64 measurements of $\phi_\mathrm{1}$ and 57 of
$\phi_\mathrm{2}$.  Figure~\ref{fig:phi} shows the histograms of all
$\phi_\mathrm{1}$ and $\phi_\mathrm{2}$ measurements collected into
phase bins of width 0.005. The mean values for all five observing
seasons combined are $\langle \phi_\mathrm{1}\rangle\!=\!0.361$ and
$\langle \phi_\mathrm{2}\rangle\!=\!0.189$, with standard deviations
of the distributions $\sigma_\mathrm{1}\!=\!0.009$ and
$\sigma_\mathrm{2}\!=\!0.005$, respectively. With exposure times of
30\,s or 60\,s in the individual light curves (not considered
separately) and an orbital period of 5388\,s, the minimal timing error,
taken as 0.5 bin widths, is between 0.003 and 0.006 phase units. Noise
in the light curves may increase this uncertainty. The standard
deviation of $\phi_\mathrm{2}$ is consistent with being entirely due
to measurement errors. The $\phi_\mathrm{1}$ distribution is wider,
indicating some intrinsic scatter, possibly caused by emission from
the second pole. In the statistical analysis, the isolated value
$\phi_\mathrm{1}\!=\!0.414$ in Fig.~\ref{fig:phi} (upper left panel)
was excluded. It was observed on 12 February 2013, close in time to
the abnormal event of Fig.~\ref{fig:lc} (bottom panel).

\begin{figure}[t]
\includegraphics[bb=131 47 488 710,height=90mm,angle=-90,clip]{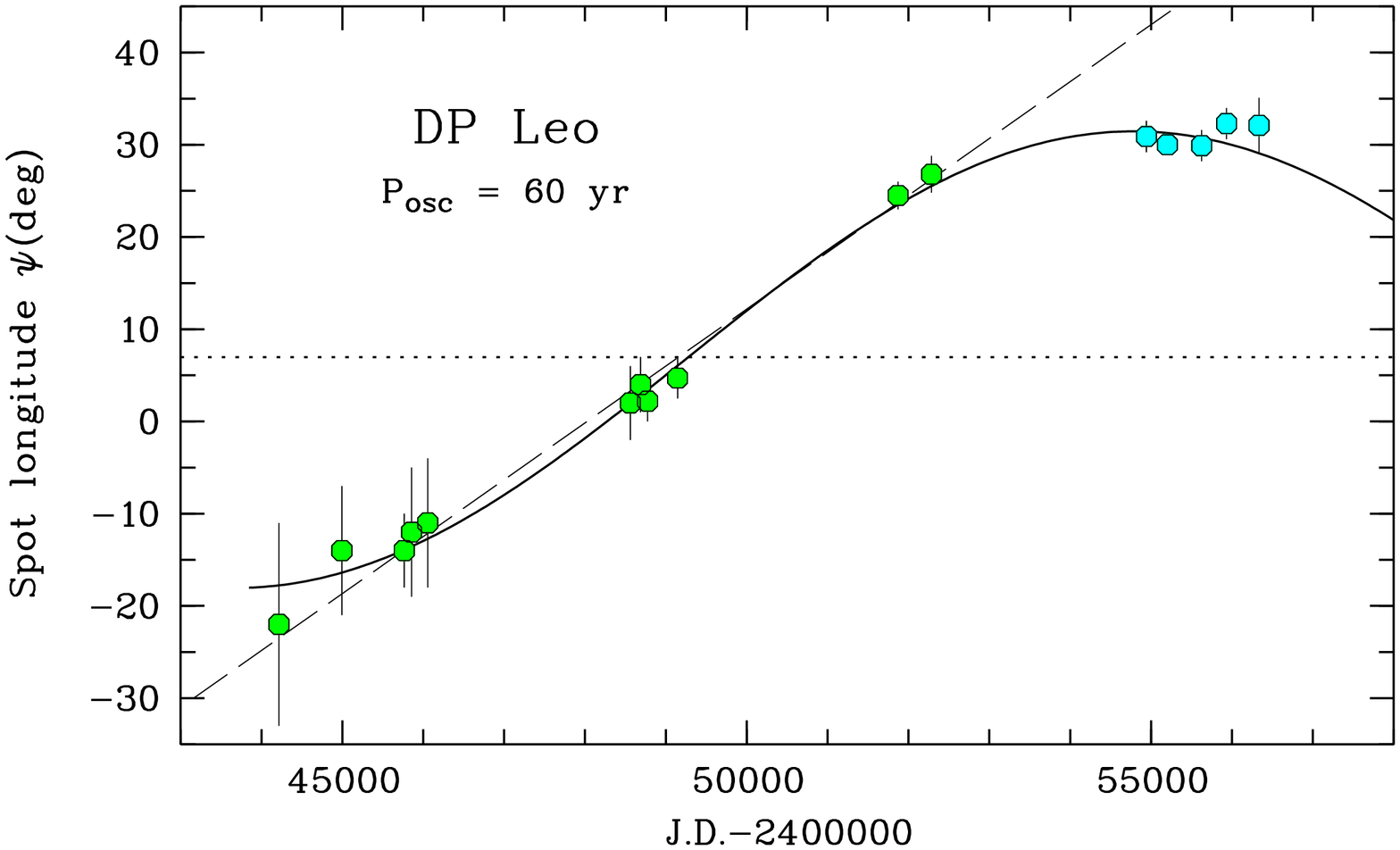}	
\includegraphics[bb=340 47 540 710,height=90mm,angle=-90,clip]{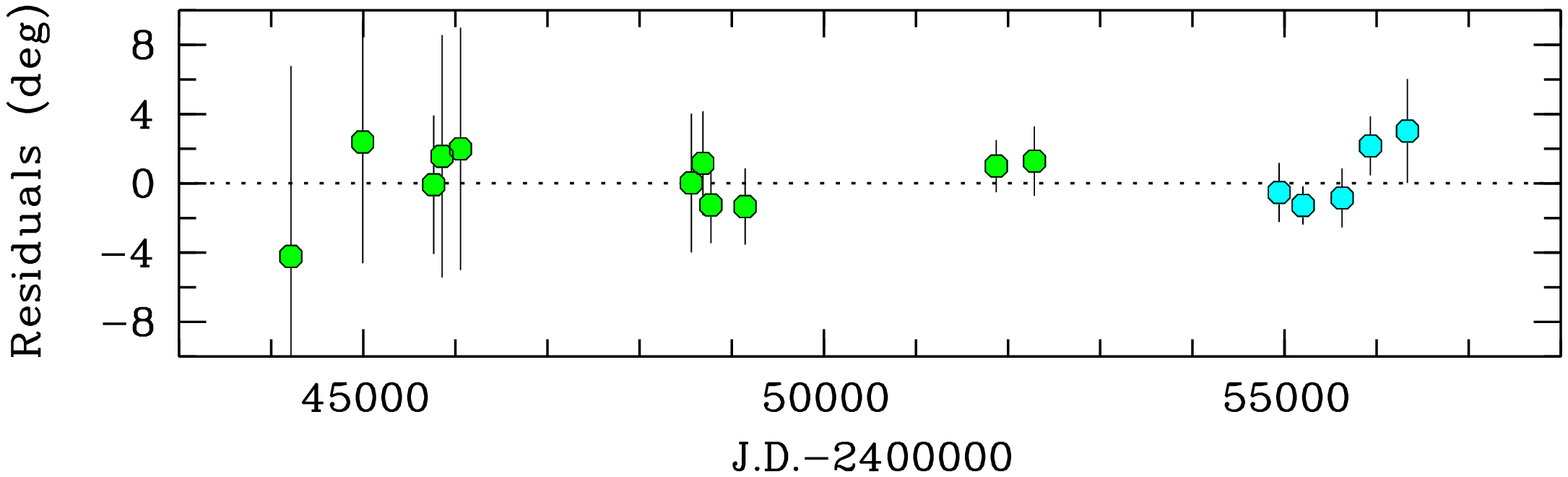}	
\caption[chart]{\emph{Top: } Time dependence of the longitude of the
  prime-pole accretion spot. \emph{Bottom: } Residuals from the
  sinusoidal fit. }
\label{fig:long}
\end{figure}

\subsection{Temporal variation of the longitude of the prime accretion spot}
The mean values of $\phi_\mathrm{1}$ and $\phi_\mathrm{2}$ for the
five individual observing seasons and for all seasons combined are
listed in Table~\ref{tab:phi} along with the corresponding longitudes
of the prime-pole accretion spot, $\psi\!=\!(\phi_2-\phi_1)\times
180\degr$. This longitude is essentially constant over the 2009--2013
period, despite variations in the accretion rate $\dot M$. This result
indicates that wandering of the accretion spot across the surface of
the white dwarf as a response to $\dot M$ variations are unimportant
in the present context. That $\dot M$ does not significantly affect
the spot longitude was previously noted for instance by
\citet{schwopeetal02}.  For completeness, we summarize the previously
published spot longitudes in
Table~\ref{tab:published}. Fig.~\ref{fig:long} shows the time
variation of the spot longitude. The published values are displayed as
green circles, the new ones as cyan-blue filled circles. Our
measurements deviate from the extrapolation of the pre-2003 linear
increase (dashed line) by highly significant $\sim\!11$\degr\ in 2009
and $\sim\!18$\degr\ in 2013. We conclude that the secular progression
of the spot longitude has halted, and that the combined data are
inconsistent with the notion mv of an asynchronously rotating white dwarf
(dashed line).

\begin{table}[t]
\begin{flushleft}
\caption{Previously published prime-pole spot longitudes $\psi$.}
\begin{tabular}{l@{\hspace{2mm}}c@{\hspace{2mm}}c@{\hspace{2mm}}c@{\hspace{2mm}}l}
\hline\hline \\[-1ex]
  Year & Mean JD  &  Band  & $\psi$   & Reference  \\
       & 2450000+ &        & (degree) & \\[0.5ex]
\hline\\
1979     & 44214.62 & X   & $-22.0\pm11.0$ & Biermann et al. (1985) \\
1982     & 44993.96 & Opt & \hspace{-1.5mm}$-14.0\pm7.0$  &  \\
1984     & 45763.83 & Opt & \hspace{-1.5mm}$-14.0\pm4.0$  &  \\
1984     & 45854.88 & X   & \hspace{-1.5mm}$-12.0\pm7.0$  & Robinson \& C\'{o}rdova (1984) \\
1985     & 46054.94 & X   & \hspace{-1.5mm}$-11.0\pm7.0$  &  \\
1991     & 48560.56 & FUV & \hspace{1.5mm}$3.0\pm4.0$    & Stockman et al. (1994) \\
1992     & 48686.50 & Opt & \hspace{1.5mm}$4.0\pm3.0$    & Bailey et al. (1993)\\
1992     & 48773.22 & X   & \hspace{1.5mm}$2.2\pm2.2$    & Schwope et al. (2002) \\
1993     & 49144.09 & X   & \hspace{1.5mm}$4.7\pm2.2$    &  \\
2000     & 51870.84 & X   & $24.1\pm2.2$   &  \\
2000     & 51870.84 & X   & $25.0\pm1.0$   & Pandel et al. (2002) \\
2002     & 52284.70 & Opt & $26.6\pm3.0$   & Schwope et al. (2002) \\ 
&&& & and this work\\[1.0ex] 
\hline\\[-1ex]
\end{tabular}
\label{tab:published}
\end{flushleft}
\end{table}

We have tested the alternative model of an oscillation of
the magnetic axis of the white dwarf about an equilibrium position by
fitting a sinusoid to the data (solid curve). The fit has a
$\,\chi^2\!=\!6.8$ for 12 degrees of freedom and yields a period of
$60\pm 9$\,yr, an amplitude of $25\degr\pm 3\degr$, and an equilibrium
orientation of $7\degr\pm3\degr$. The fit plausibly explains all
available data on the temporal variation of the spot longitude.

\section{Discussion and conclusion}
\label{sec:disc}

We have, for the first time, presented evidence for the oscillation or
libration of the magnetic axis of the white dwarf in a polar that
results from the combined action of the accretion and magnetic
torques. The theory of the magnetostatic interaction of the magnetic
moments of white dwarf and the secondary star predicts an oscillation
period on the order of a few decades and an amplitude of a few tens of
degrees about an equilibrium longitude of the accreting pole that
slightly leads the line connecting the stars
\citep{kingetal90,kingwhitehurst91,wickramasinghewu91,wuwickramasinghe93,campbellschwope99}.
A period of about 60\,yr, an amplitude of 25\degr, and an equilibrium
longitude of about 7\degr\ are entirely consistent with
theory. \citet{schwopeetal02} had argued that with increasing $\psi$
the second pole becomes more easily accessible for the accretion
stream. The enhanced cyclotron emission detected by us from the second
pole is in line with this argument. If the magnetic axis of the white
dwarf in \dpleo\ is, in fact, turning back, we predict that
second-pole emission again becomes rare in the years to come.  We
caution, however, that the data cover only one half of the implied
oscillation period, and we do not claim to have presented final proof
of the proposed oscillation. Proving the periodicity and the existence
of a secularly stable equilibrium position requires several more decades
of monitoring of \dpleo.

\begin{acknowledgements}
  Our data were obtained with the MONET/North telescope of the
  MOnitoring NEtwork of Telescopes, funded by the Alfried Krupp von
  Bohlen und Halbach Foundation, Essen, and operated by the
  Georg-August-Universit\"at G\"ottingen, the McDonald Observatory of
  the University of Texas at Austin, and the South African
  Astronomical Observatory.  The ``Astronomie \& Internet'' program of
  the Foundation and the MONET consortium provide observation time to
  astronomical projects in high-schools worldwide. Part of the early
  data were obtained in collaboration with Jens Diese and
  his high-school students at the Max-Planck-Gymnasium, G\"ottingen,
  Germany.
\end{acknowledgements}

\bibliographystyle{aa}

\end{document}